\documentclass[a4paper, amsfonts, amssymb, amsmath, reprint, showkeys, nofootinbib, twoside,superscriptaddress]{revtex4-1}
\usepackage[english]{babel}
\usepackage[utf8]{inputenc}
\usepackage{amsthm}
\usepackage{amssymb}
\usepackage{mathtools}
\usepackage{physics}
\usepackage{xcolor}
\usepackage{graphicx}
\usepackage[left=23mm,right=13mm,top=35mm,columnsep=15pt]{geometry} 
\usepackage{adjustbox}
\usepackage{placeins}
\usepackage[T1]{fontenc}
\usepackage{lipsum}
\usepackage{csquotes}
\usepackage[pdftex, pdftitle={Article}, pdfauthor={Author}]{hyperref} % For hyperlinks in the PDF
\hypersetup{
    colorlinks=true,
    linkcolor=blue,
    filecolor=magenta,      
    urlcolor=cyan,
}

\usepackage[normalem]{ulem}

\begin{document}
\title{Confusion-driven machine learning of structural phases\\ of a flexible, magnetic Stockmayer polymer}

\author{Dilina Perera}
    \email[Correspondence email address: ]{dilina.perera@ung.edu}
    \affiliation{Department of Physics and Astronomy, University of North Georgia, Dahlonega, GA 30597, U.S.A.}
\author{Samuel McAllister}
    \thanks{Current address: University of Alabama at Birmingham, Birmingham, AL 35294, U.S.A.}
    \affiliation{Department of Physics and Astronomy, University of North Georgia, Dahlonega, GA 30597, U.S.A.}
\author{Joan Josep Cerd\`{a}}
\affiliation{Departament de F\'isica UIB i Institut d'Aplicacions Computacionals de Codi Comunitari (IAC3), Campus UIB, E-07122 Palma de Mallorca, Spain}
\author{Thomas Vogel}
    \email[Correspondence email address: ]{thomasvogel@lanl.gov}
        \affiliation{Computer, Computational, and Statistical Sciences Division, Los Alamos National Laboratory, Los Alamos, NM 87545, U.S.A.}

\date{\today} % Leave empty to omit a date

\begin{abstract}
We use a semi-supervised, neural-network based machine learning technique, the confusion method, to investigate structural transitions in magnetic polymers, which we model as chains of magnetic colloidal nanoparticles characterized by dipole-dipole and Lennard-Jones interactions. As input for the neural network we use the particle positions and magnetic dipole moments of equilibrium polymer configurations, which we generate via replica-exchange Wang--Landau simulations. We demonstrate that by measuring the classification accuracy of neural networks, we can effectively identify transition points between multiple structural phases without any prior knowledge of their existence or location. We corroborate our findings by investigating relevant, conventional order parameters. Our study furthermore examines previously unexplored low-temperature regions of the phase diagram, where we find new structural transitions between highly ordered helicoidal polymer configurations.
\end{abstract}

\maketitle

\section{Introduction} \label{sec:intro}

In this work we use machine-learning techniques to study conformational phases of magnetic colloidal polymers, also known as magnetic filaments. The synthesis of magnetic colloidal polymers has advanced significantly in the last two decades. Today, the bonds between neighboring colloids in the chain can be formed in many different ways~\cite{2003-goubault,2005-goubault,2007-martinez-pedrero,2007-evans,2008-martinez-pedrero,2008-liu,2005-harpreet,2008-benkoski,2009-zhou,2011-benkoski,2011-wang,2012-sarkar,2012-breidenich,2013-busseron,2014-byrom,2014-lawrence}, allowing the production of colloidal polymers with different lengths, degrees of flexibility, interaction between non-adjacent colloids, and magnetic properties. The growing interest in suspensions of magnetic filaments is due to their potential applications, such as microfluids, improved substitutes for current ferrofluids, elements for magnetic information storage, as well as pressure and chemical nanosensors, to mention just a few~\cite{2011-wang}. Inverse magnetic filaments have been used in the field of the assembly of non-permanent photonic crystals~\cite{2012-liu}. 

Analytical approaches~\cite{2004-shcherbakov,2008-erglis-mh,2005-cebers,2004-cebers,2009-belovs-pre,2009-belovs-jpa,2009-snezhko,2014-morozov-ns} and experiments~\cite{2008-erglis-jpcm,2008-erglis-mh,2003-biswal,2011-benkoski} have been shown to be very useful in the determination of the properties of systems formed by colloidal para- and super-paramagnetic polymers. In the case of ferromagnetic colloids, several works have been devoted to analytical~\cite{1997-pitard,2011-phatak}, experimental~\cite{2005-singh,2011-trevisan,2006-korth}, and numerical studies of these kind of systems~\cite{2005-cerda-jcp}. In numerical simulations different setups have been tested in the past: colloidal polymer systems at infinite dilution~\cite{2013-cerda,2016-cerda,2011-sanchez,2013-sanchez,2015-pshenichnikov,2016-luesebrink}, finite concentration suspensions~\cite{2016-stankovic,2024-cerda,2002-morozov}, and subsequently ensembles of colloidal polymers forming arrangements known as polymer brushes~\cite{2015-sanchez-mm,2019-cerda,2021-cerda,2015-sanchez-b}.

As most of the technological applications are related to the magneto-responsive control of these materials, it is crucial to have a good understanding of the different types of equilibrium configurations that isolated colloidal polymers exhibit. Two earlier works by Cerd\`{a} \textit{et al.}~\cite{2013-cerda,2016-cerda} have established the conformational phase diagram for ferromagnetic colloidal polymers at zero field and under an applied external field, respectively. These studies, investigating Stockmayer colloidal polymers~\cite{stockmayer1,stockmayer2}, combined Lennard-Jones (LJ) potentials with the interaction of magnetic dipoles located at the center of the colloids. They examined in detail the dependence of the chain conformations on the relative strength of the LJ compared to the magnetic interaction. This ratio can be characterized via the dimensionless parameter
\begin{equation} \label{eta-definition}
\eta \equiv\frac{\epsilon \sigma^3}{\mu^2}\,,
\end{equation}
where $\epsilon$ and $\sigma$ set the strength and range of an attractive
Lennard-Jones type potential between two colloids and $\mu^2$ defines the
magnetic dipole strength.\footnote{$\eta$ effectively compares the attractive and magnetic
energies in a situation where two
ferromagnetic colloids are in close contact, with their dipoles
oriented in a nose-tail conformation, that is, for two dipoles $i$ and
$j$, $\vec{\mu}_i \cdot \vec{\mu}_j = \mu^2$ and $\vec{\mu}_i \cdot \vec{r}_{ij} = \vec{\mu}_j \cdot \vec{r}_{ij} = \pm \mu r_{ij}$, which minimizes the magnetic
energy. See below.}

\begin{figure}
    \centering
    \includegraphics[width=\columnwidth]{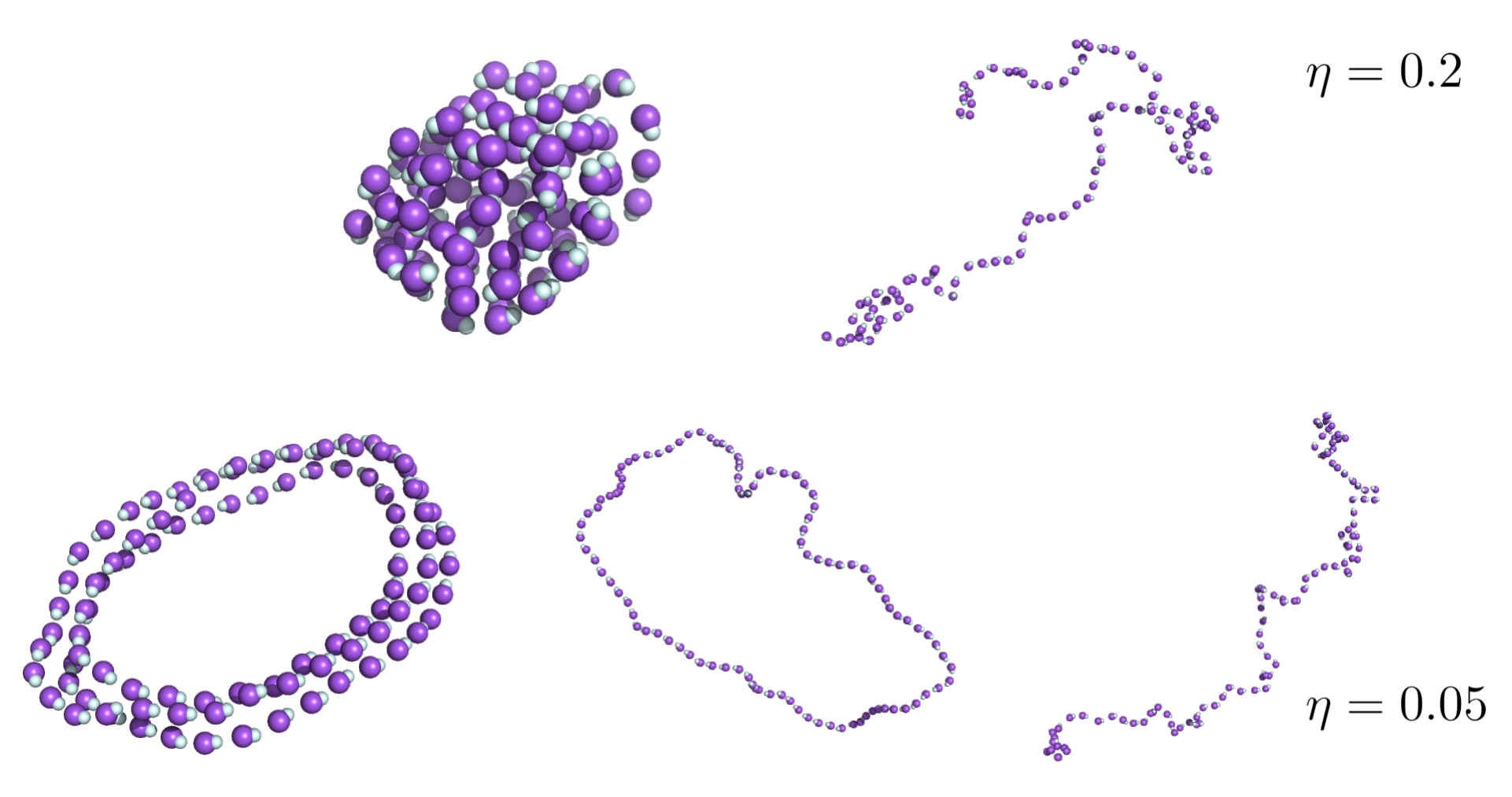}
    \caption{Configurations of a magnetic Stockmayer polymer at $\eta=0.2$ (top row) and $\eta=0.05$ (bottom row), all at $\mu^2=5.0$, at different temperatures (increasing from left to right). For $\eta=0.2$ the configurations illustrate the transition from compact, disordered structures at low temperatures to extended, open ones at higher temperatures. At $\eta=0.05$, the magnetic interaction is stronger relative to the Lennard-Jones attraction, leading to the formation of helicoidal, closed structures with aligned magnetic dipoles (white dots) at low temperatures.}
    \label{fig:joan_configs}
\end{figure}

\begin{figure}
    \centering
    \includegraphics[width=.923\columnwidth]{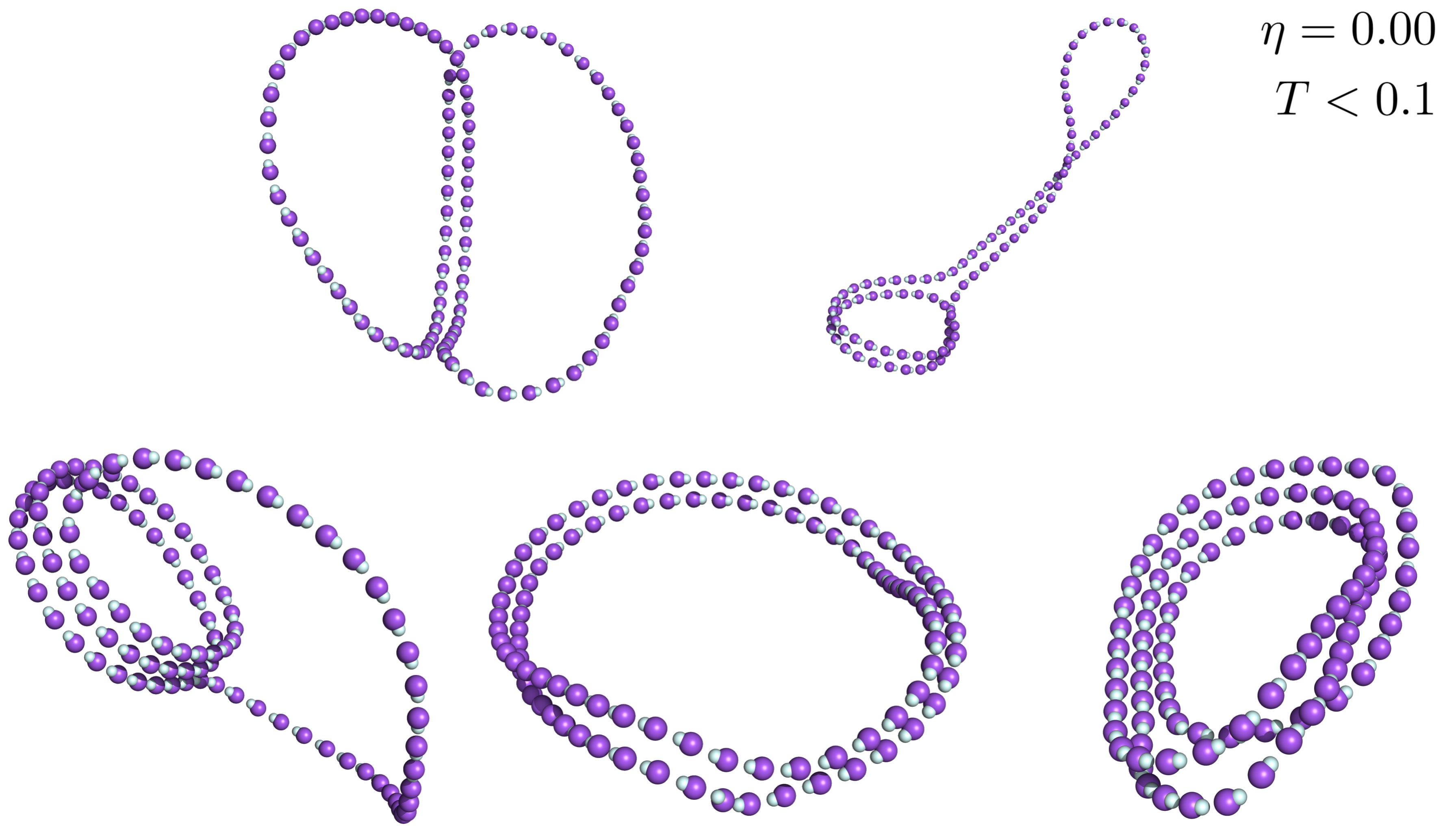}
    \caption{Configurations of a magnetic Stockmayer polymer at $\eta=0$ (that is, there is no attractive interaction between the colloids, just a magnetic dipole-dipole interaction) at very low temperatures. These structured are located in previously unexplored regions in the phase diagram.}
    \label{fig:eta0_configs}
\end{figure}

In~\cite{2013-cerda,2016-cerda}, the authors focused on polymers with $0\leq \eta \leq 0.2$ at temperature $T\gtrsim 0.3$, finding various classes of chain conformations; see Fig.~\ref{fig:joan_configs} for examples. The temperature region at $T<0.3$ was not addressed and remained unexplored, as Langevin simulations are difficult to equilibrate at such low temperatures. In this work, we aim at exploring this region and show how machine-learning methods can be used to construct the corresponding phase diagram. As a preview, Fig.~\ref{fig:eta0_configs} shows magnetic polymer structures that we find at very low temperatures, $T<0.1$, for $\eta=0$, that is, where the energy is determined only by the magnetic dipole-dipole interaction. We found a great variety of different looped structures in this previously unexplored region.

The use of machine learning (ML) in many branches of physics~\cite{2021-tanaka,2022-knecht,2023-choudhary,2024-brunton,2024-beaucage} and other scientific disciplines~\cite{Schmidt2019npj,Choudhary2022npj,2024-sridharan} has received considerable attention in the last decade.  For example, ML has been applied to the description of many-body systems~\cite{2024-sammueller}, it helped to determine the transition temperatures of glass in vitrimers~\cite{2024-karatrantos}, it was used to study colloidal self-assembly~\cite{2021-dijkstra}, to determine phase diagrams~\cite{2024-noordhoek,2023-chew}, applied in rheological research~\cite{2024-miyamoto} and membrane systems~\cite{2022-sawade}, in the study of tribology~\cite{2023-sose}, the determination of particle-particle interaction potentials~\cite{2024-matin,2023-kulichenko,2023-yu,2022-wen,2022-casadio,2019-deringer}, or the description of crystal nucleation~\cite{2023-beyerle}. For more comprehensive reviews discussing machine learning in condensed matter physics and polymer research, see~\cite{2025-stenzel,2024-zhang,2024-barrat,2023-li,2021-clegg,2021-bedolla,2019-jackson}, for example.

    ML methods fall into two broad categories: supervised learning, which requires the model to be trained using input samples with known categories (class ``labels''), and unsupervised learning, which does not require such prior training with labeled data. In the study of phase transitions and critical phenomena, both supervised~\cite{Carrasquilla2017np, Schindler2017prb, zhang2017prl, Chng2017prx, Ponte2017prb, Li2018Ann, ZhangP2018prl} and unsupervised methods (particularly dimensionality reduction techniques)~\cite{Wang2016prb, Wang2017prb, Hu2017pre, Costa2017prb, Wetzel2017pre, Chng2018pre} have been extensively explored. Dimensionality reduction methods such as principal component analysis, multidimensional scaling, and autoencoders yield two-dimensional representations of the configuration space, enabling one to visually distinguish different thermodynamic and structural phases. In systems with well-defined order parameters (e.g., the Ising model or the XY model), the dominant principal components or the latent variables have demonstrated a linear correlation with these parameters~\cite{Hu2017pre, Wetzel2017pre}, while in systems lacking such order parameters, they can effectively serve as proxies for characterizing phase transitions~\cite{Parker2022pre}.

    In the domain of supervised learning, neural networks (NNs) trained on labeled configurations have been used to identify the transition point between two distinct phases by examining the intersections of average NN outputs as functions of temperature, energy, or a model parameter~\cite{Carrasquilla2017np, Wei2017pre, Parker2022pre}. Through finite-size scaling analyses, one can then estimate the critical temperature and exponents~\cite{Carrasquilla2017np}. An inherent limitation of this approach is the requirement for a labeled data set to train the NN. For systems such as the Ising model or the XY model with well-known phases, configurations can be categorized into distinct phases based on an order parameter. However, for more complex systems with little to no prior knowledge of the phases the system may exhibit, this labeling process can become significantly more challenging, or even impossible.  

    An alternative approach to detect phase transitions using NNs was proposed by van Nieuwenburg \textit{et al.}~\cite{Nieuwenburg2017np}. Commonly known as the ``Confusion Method'', it does not require that the true labels of the training samples be known beforehand. Instead, configurations are labeled on the basis of a hypothetical transition point, and changes in the network performance are evaluated as this hypothetical transition point is systematically changed. Consequently, the method does not require any prior knowledge about the existence or absence of transitions. The confusion method has been successfully applied to detect transitions in a wide array of systems, including classical spin models~\cite{Nieuwenburg2017np, Beach2018prb, Lee2019pre}, polymers~\cite{Xu2019pre, Parker2022pre}, quantum systems~\cite{Nieuwenburg2017np, Zvyagintseva2022cp, Gavreev2022njp}, and complex networks~\cite{Ni2019pre}. Furthermore, the method has been shown to also be effective in simultaneously detecting multiple transitions~\cite{Lee2019pre}.
    
    In this paper, we apply the confusion scheme to determine boundaries between multiple phases of a flexible Stockmayer polymer.  To train the NN we use two types of data: the position coordinates and the magnetic dipole moments of the equilibrium polymer configurations, both of which were found to yield qualitatively similar results. The main aim of the present work is to test the efficiency and performance of the method against already known results~\cite{2013-cerda}. As we will show, the confusion method constitutes an excellent tool with clear advantages over conventional methods to determine the conformational phase diagram of polymer filaments in the limit of infinite dilution.
    Although the full determination of the phase diagram for these colloidal magnetic polymers is out of the scope of the present work, we did identify additional conformational transition in the low-energy regime which has not been explored previously.

    The manuscript is organized in the following way. In Sec.~\ref{sec:model}, we introduce the numerical model we use, in Sec.~\ref{sec:methods} we provide the details of our methods. The main findings are presented in Sec.~\ref{sec:results}, and conclusions are given in Sec.~\ref{sec:summary}.
\section{Model} \label{sec:model}

The Stockmayer polymers studied in this work are composed of $N$ ferromagnetic particles (monomers), represented as beads permanently linked by bonds between their surfaces. As done previously~\cite{2013-cerda} we use reduced units: the index $e$ on a dimensional quantity denotes an experimental value, whereas the absence of such an index means that the quantity is expressed in reduced units. For example, all lengths in the system are related to units of monomer diameter, that is, $l=l_e/\sigma_e$, so that the reduced diameter of a polymer monomer is $\sigma=1$. In the case of the energies, if we take $\epsilon_e$ as the characteristic experimental energy scale, then any energy $U_e$ will be expressed in reduced units as $U=U_e/\epsilon_e$.

 To mimic the steric effects between two monomers $i$ and $j$ we use the Weeks--Chandler--Andersen potential~\cite{1971-weeks}, a truncated and shifted Lennard-Jones (LJ) interaction that is purely repulsive: 
\begin{align}
      \nonumber
      & U_{\mathrm{rep}} (r) =\\ & \left\{ \begin{array}{ll} 
        U_{\mathrm{LJ}}(r, 1, 1)-U_{\mathrm{LJ}}(r_{\mathrm{cut}}, 1, 1)\,,&
            \mbox{for $ r < r_{\mathrm{cut}}=2^{1/6}$ }  \\ 
        0\,,& \mbox{for $ r  \geq  r_{\mathrm{cut}}=2^{1/6}$} 
\end{array} \right.,  \label{WCA}
\end{align}
where $r=| \vec{r}_i-\vec{r}_j |$ is the reduced distance between the centers of the particles, and
\begin{equation}\label{eq:LJ}
U_{\mathrm{LJ}}(r, \epsilon, \sigma) \equiv 4 \epsilon \left[ (\sigma/r)^{12}-(\sigma/r)^6 \right]\,.
\end{equation}

In addition to steric repulsion, we introduce a central attraction between monomers to simulate the effects of poor solvents (Stockmayer polymers are often referred to as ``sticky chains''). This interaction is also modeled as a truncated and shifted Lennard-Jones potential with a cutoff $r_{\mathrm{cut}}=2.5\sigma$ (with $\sigma=1$):
\begin{align}
      \nonumber
      & U_{\mathrm{att}} (r,\epsilon) =\\ & \left\{ \begin{array}{ll} 
      U_{\mathrm{LJ}}(r,\epsilon,1) - U_{\mathrm{LJ}}(r_{\mathrm{cut}},\epsilon,1)\,, &     \mbox{for $ r < r_{\mathrm{cut}}=2.5$ }  \\ 
      0\,,& \mbox{for $ r  \geq r_{\mathrm{cut}}=2.5$} 
\end{array} \right..  \label{sticky}
   \end{align}
By varying $\epsilon$ we control the parameter $\eta$, see Eq.~(\ref{eta-definition}).

\begin{figure}[b]
\includegraphics[width=.71\columnwidth]{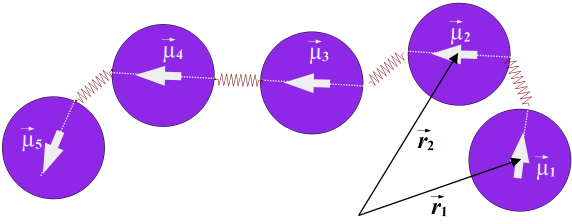}	
\caption{\label{fig:model}Schematic representation of the bead--spring model with magnetic dipoles and bond--dipole coupling. Note that the links between adjacent surfaces do not only constrain the inter-particle distance but also the relative orientation of the dipoles. The anchoring points for each colloid are given by $\pm(\sigma/2)\hat{u}$  where  $\hat{u}=\vec{\mu}/\mu$ with respect to the centre of the colloids. The bonds themselves are modeled as harmonic springs.\vspace{-.4em}}
\end{figure}
To capture magnetic effects, we assume the point dipole approximation and assign such a magnetic dipole $\vec{\mu}_e$ to the center of each colloidal monomer; see Fig.~\ref{fig:model}. The dipole is fixed within the particle body frame, that is, it can only rotate if the whole monomer performs a rotation. The modulus of the magnetic moments in the reduced units is calculated as $\mu^2 = \mu^2_e /(4 \pi \mu_{0}\, \sigma^3_e\epsilon_e)$, where $\mu_{0}$ is the vacuum magnetic permittivity. 
The magnetic dipole--dipole interaction is then expressed as:
\begin{equation}
\label{dipdip} 
U_{\mathrm{dip}}(\vec{r}_{ij},\vec{\mu}_i,\vec{\mu}_j) =
\frac{\vec{\mu}_i \cdot \vec{\mu}_j }{r^3} - 
\frac{3\,[\vec{\mu}_i \cdot \vec{r}_{ij}][\vec{\mu}_j \cdot \vec{r}_{ij} ]}{r^5}\,,
\end{equation}
where $\vec{r}_{ij} = \vec{r}_i - \vec{r}_j$ is the displacement vector between particles
$i$ and $j$. Note that in the previous formula we assume Gaussian units. Reasonable values of $\mu=|\vec{\mu}|$ depend on the composition and size of the nanoparticles, but typically do not exceed values of $\mu \approx 10$ for common ferrofluids. Note that we do not introduce a Zeeman energy here, as previously done~\cite{2016-cerda}, since we do not consider external fields in this work.

The bond between consecutive monomers is modeled using a spring that connects to anchoring points at the surface of the particles, see also~\cite{2013-cerda,2016-cerda,2015-sanchez}. 
We use these surface-to-surface links to mimic the experimental way of crosslinking ferromagnetic colloids by means of molecules that are end-grafted to the surface of neighbor particles in the chain~\cite{2014-byrom,2014-hu-ns,2015-fratila}. The rotation of the dipole moments away from the filament backbone is hence penalized by the mechanical stretching of the springs, favoring head-to-tail conformations. 
The use of center-to-center potentials to mimic the bonds between adjacent monomers in the chain is not advisable in this case, as particles could then freely rotate while having their positions locked. That could easily lead to unnatural conformations in which, for instance, two consecutive monomers in the chain are side by side, but with their dipoles pointing in opposite directions. In order to avoid such structures, the potential we use to model such surface-to-surface links reads
 \begin{align}
     \nonumber
     &U_{\mathrm{bond}}(\vec{r}_i,\vec{r}_{i+1},\hat{u}_i,\hat{u}_{i+1}) \\ &\quad= 
     \frac{1}{2} K_{\mathrm{S}} \left(\vec{r}_i - \vec{r}_{i+1} - \left(\hat{u}_i+\hat{u}_{i+1}\right)\, \frac{\sigma}{2} \right)^2,
     \label{SOSpotencial}
  \end{align}
where $\hat{u}_i=\vec{\mu}_i/\mu_i$ is the unit vector in the direction of the magnetic moment, pointing at the anchoring surface point of bead $i$, and $\vec{r}_i$ and $\vec{r}_{i+1}$ are the position vectors to the centers of two linked monomers, as shown in Fig.~\ref{fig:model}. The anchoring points for the springs on monomer $i$ are collinear and located at $\vec{r}_i\pm \vec{l}_i$ with $\vec{l}_i=\hat{u}_i\,\sigma/2$. The constant of the bond interaction is set to $K_{\mathrm{S}}=30$, which results in the typical bond length being approximately within the range $[0.98,1.1]$. The alignment of the dipoles with the anchoring points at the surface of the particles is a model used earlier by Cerd\`{a} \textit{et al.}~\cite{2013-cerda,2016-cerda}. We adhere to it to reliably test to which extent our method is able to reproduce the conformational phase diagram. The rationale behind assuming the alignment of dipoles and anchoring points is that in many experimental works colloidal polymers follow that kind of linking pattern~\cite{2011-trevisan,2006-korth,2009-zhou,2014-hu-ns,2015-fratila}. It should be noted, however, that numerical method we use is completely generic and can be applied to any other polymer chain models just as well. 

The total energy in our model is then:
\begin{equation}
    U_{\mathrm{tot}}=\sum_{i,j;\,i\neq j}\left(U_{\mathrm{rep}}+U_{\mathrm{att}}+U_{\mathrm{dip}}\right)+\sum_{i<N-1}U_{\mathrm{bond}}\,,
\end{equation}
where $N$ is the total number of magnetic monomers.

Finally, we use the reduced temperature $T=kT_e/\epsilon_e$, where $T_e$ is the actual temperature of the sample and $k$ is the Boltzmann constant.

\section{Methods} \label{sec:methods}

\subsection{Generating equilibrium polymer configurations} \label{sec:methods.rewl}

\begin{figure}[b]
\includegraphics[viewport=14 13 906 578, width=\columnwidth, clip]{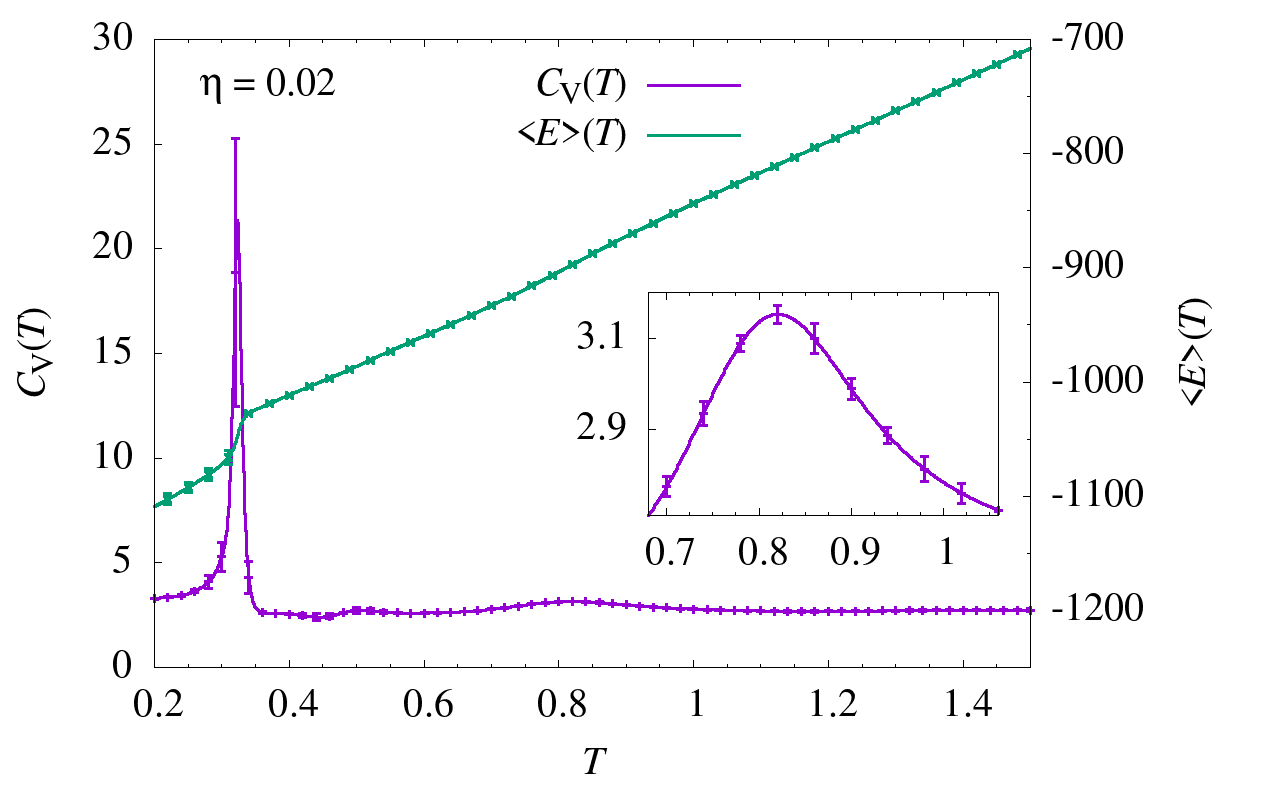}
\includegraphics[viewport=14 13 906 578, width=\columnwidth, clip]{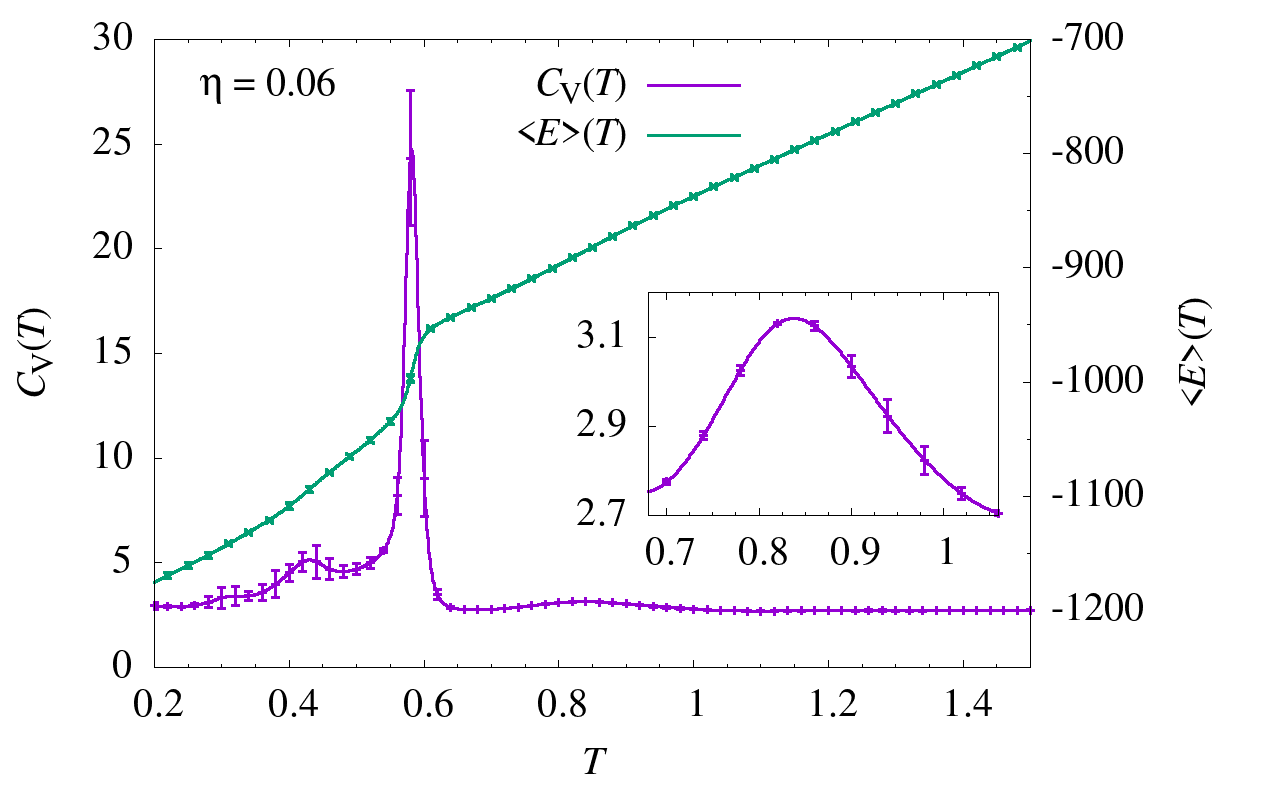}
\caption{The heat capacity, $C_\mathrm{V}$, and canonical average energy, $\langle E\rangle$, as functions of the canonical temperature for polymers of length $N=100$ after Wang--Landau runs have converged. At this time we start collecting configurations for the machine-learning analysis. Top: $\eta=0.02$, bottom: $\eta=0.06$. The inset emphasizes the heat capacity in an interval around the weaker peak at $T\approx 0.83$.\label{fig:thermo}\vspace{-.4em}}
\end{figure}

We generate polymer configurations using the replica-exchange Wang--Landau (REWL) method~\cite{2013-vogel-prl,2014-vogel-pre,2018-vogel-jpcs}, splitting the whole energy range in 19 overlapping intervals of equal size. The overlap between two neighboring energy intervals is 75\%. For each value of $\eta$ we first search for a lower bound of the energy range so that equilibrium configurations down to temperatures of $T\approx0.2$ are sampled sufficiently well. To equilibrate the system, we use a common flatness criterion and update scheme for the Wang--Landau modification factor~\cite{2001-wang-prl,2001-wang-pre}, that is, until the Wang-Landau histogram is reasonably flat at a modification factor of $f<1.0000001$. From the estimator of the density of states, $g(E)$, that we obtain from these runs we calculate the canonical average of the energy, $\langle E \rangle_{\textrm{can}}(T)$ and the heat capacity, $C_\mathrm{V}(T)$~\cite{2014-li-jpcs}. This canonical thermodynamic analysis is shown in Fig.~\ref{fig:thermo} for magnetic polymers of length $N=100$. With the system then equilibrated, that is, when randomly walking in energy space, we perform a collection run where we record polymer configurations from all energy bins. To limit correlations between the configurations we store to train the NN, we enforce a minimum distance in energy space for the Monte Carlo process to advance before recording another configuration. In total, we collect about $500\,000$ configurations, equally distributed over 200 collection bins that cover the whole energy range, for each value of $\eta$. These configurations will serve as the input data for the neural networks we employ for analysis.

\subsection{Neural networks}
A neural network (NN) is a flexible machine learning model that can be used in both supervised and unsupervised settings to learn complex non-linear relationships within data. It consists of simple information-processing units called neurons, which are interconnected according to a specified network architecture. The most common and simplest of these architectures is the layered structure known as a feed-forward network, comprising an input layer, hidden layers, and an output layer. Each neuron receives multiple inputs and produces a single output by calculating a weighted sum of the inputs and then applying a nonlinear function, known as an activation function. This output is then passed as input to the neurons in the next layer. 

During the training phase, a NN is provided with data (such as images or polymer configurations) and corresponding ``labels'', which represent the correct classification or prediction for each input sample. The model iteratively adjusts its internal parameters (i.e., weights of the neurons and biases) to minimize the difference between the true labels and the model's predictions through a process known as backpropagation.
This training enables the network to capture underlying patterns that associate the input data distribution with the true labels. Once trained, a NN can predict labels for previously unseen input samples.

When using NNs to detect structural transitions, we provide equilibrium polymer configurations generated using the REWL method as input data. Although both the positions $\{\vec{r}_i\}$ and magnetic dipoles $\{\vec{\mu}_i\}$ of the monomers are required to fully describe a polymer configuration, we find it often sufficient to use either the positions $\{\vec{r}_i\}$ or the dipoles $\{\vec{\mu}_i\}$ as input samples to identify transitions. Our feature set is therefore $3N$-dimensional, consisting of either the spatial coordinates or the components of the dipoles of the $N$ monomers. 

\subsection{The confusion method}

Suppose the structure of the configurations depends on a parameter $E$. This parameter may represent a thermodynamic property, such as the energy of the configuration (as in our case) or the temperature, or it could be a model parameter that influences the underlying structure (see, e.g.~\cite{Parker2022pre}). Assume that there exists a point $E_c$ within an interval $[E_a, E_b]$ at which the system undergoes an abrupt structural change, so that the configurations with $E$ below and above $E_c$ have distinct structural characteristics. The confusion method seeks to locate the transition point $E_c$ by proposing trial transition points $E_c^\prime$ within the interval $[E_a, E_b]$. Labels are assigned to configurations based on $E_c^\prime$ and the NN is trained to classify these configurations according to this label assignment. For a given trial transition point $E_c^\prime$, each configuration $X_i$ with energy $E(X_i)$ is assigned a label $y_i$ as
\begin{equation}
    y_i = \begin{cases}
             0, & \text{if} \; E(X_i) \leq E_c^\prime \\
             1, & \text{otherwise}
        \end{cases}.
\end{equation}
The configurations $\{X_i\}$ and the labels $\{y_i\}$ are then split into a training and a test set. The NN is trained on the training set and then evaluated on the test set to obtain the test classification accuracy $P(E_c^\prime)$, which represents the fraction of correctly classified test configurations for the given $E_c^\prime$. This constitutes a single step in the confusion scheme. The process is then repeated for different trial transition points $E_c^\prime$ by systematically varying $E_c^\prime$ within the interval $[E_a, E_b]$. The resulting curve $P(E_c^\prime)$ versus $E_c^\prime$ will form a ``W'' shape, with the central peak located at the true transition point $E_c$. 

The characteristic W-shape in $P(E_c^\prime)$ can be explained as follows. When $E_c^\prime = E_a$, all configurations will be assigned the label ``1''. In this trivial case, the NN will learn to predict the label ``1'' for all configurations, regardless of any structural differences or similarities, thus achieving a perfect classification accuracy of 1.0. Similarly, the NN achieves perfect classification accuracy when $E_c^\prime = E_b$, where all configurations will be labeled ``0''. When $E_c^\prime = E_c$,  configurations below $E_c$ will be labeled ``0'' while those above $E_c$ will be labeled ``1''. Since the configurations below $E_c$ are structurally different from those above $E_c$, they naturally divide into two separate categories. As the assigned labels are consistent with this inherent grouping of the configurations, the NN will, in principle, not be ``confused'' and can achieve a perfect accuracy of 1.0. 

\begin{figure}[b]
\includegraphics[viewport=23 17 1056 689, width=.78\columnwidth, clip]{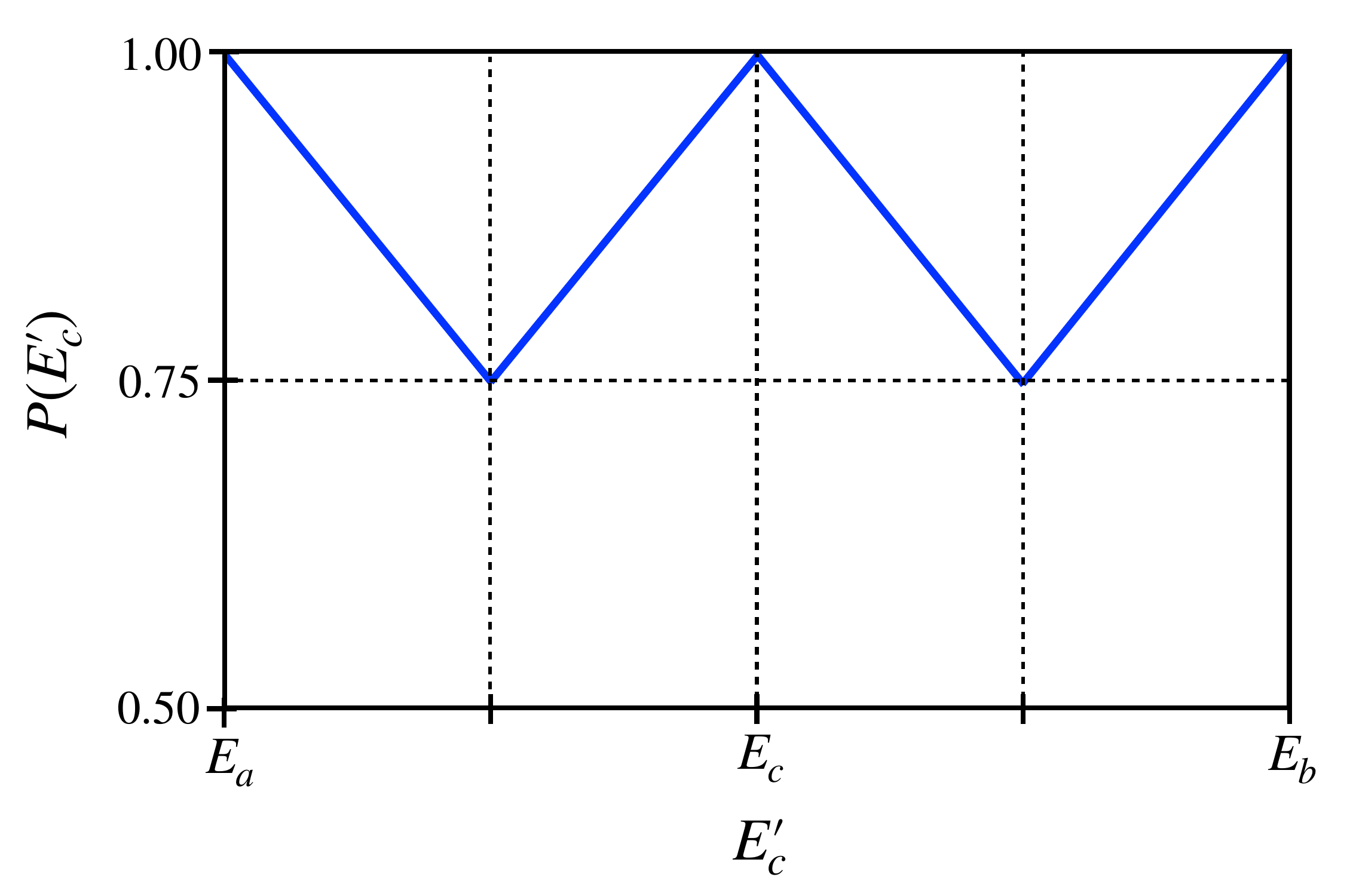}
\caption{Expected behavior of the test classification accuracy  $P(E_c^\prime)$ as a function of the trial transition point $E_c^\prime$ when the confusion method is applied over an energy range $E_c^\prime \in [E_a, E_b]$. The perfectly symmetric W-shaped profile as shown occurs only if the true transition point $E_c$ is positioned exactly at the midpoint between $E_a$ and $E_b$.\label{fig:W_shaped_profile}\vspace{-.4em}}
\end{figure}

For any other value of $E_c^\prime$ that does not coincide with $E_a$, $E_b$, or $E_c$, the NN will detect an inconsistency between the assigned labels and the natural grouping of configurations based on structural similarities and differences. A perfect classification accuracy of 1.0 is no longer possible, but the NN will learn a label assignment that minimizes misclassifications. For example, consider the case $E_a < E_c^\prime < E_c$. Within this range, we can distinguish two sub-intervals with different outcomes. If $E_a < E_c^\prime < (E_a+E_c)/2$, the NN will learn to predict the label of the majority class (i.e., ``1''), leaving a fraction of $(E_c^\prime - E_a)/(E_b - E_a)$ configurations misclassified. When $(E_a+E_c)/2 < E_c^\prime < E_c$, the NN will learn to predict ``0'' for all configurations below $E_c$ and ``1'' for all configurations above $E_c$, leaving a fraction of $(E_c - E_c^\prime)/(E_b - E_a)$ configurations misclassified. The same reasoning applies to $E_c < E_c^\prime < E_b$. Consequently, the classification accuracy can be expressed as 
\begin{equation}
    P(E_c^\prime) = 1 - \frac{\min(E_c^\prime - E_a, |E_c - E_c^\prime|, E_b - E_c^\prime)}{E_b - E_a}\,,
\end{equation}
which yields the W-shape shown in Fig.~\ref{fig:W_shaped_profile}. Note that such a perfectly symmetrical W-shape will occur only if $E_c$ is positioned exactly at the midpoint between $E_a$ and $E_b$. Otherwise, the shape will become asymmetrical with the two minima having different depths. Nevertheless, the central peak will still coincide with the true transition point $E_c$. If no transition occurs within the interval $[E_a, E_b]$, the central peak in the $P(E_c^\prime)$ curve vanishes, forming a ``V''-shape instead.

 Note that if multiple transitions are present within the interval $[E_a, E_b]$, the confusion scheme, as described, can still locate all transition points. For example, suppose there are three distinct structural phases within $[E_a, E_b]$, with transitions occurring at $E_c$ and $E_d$ (where $E_c < E_d$). When $E_c^\prime = E_c$, although the same label ``1'' is assigned to two structurally distinct groups of configurations that exist above $E_c$ (separated by the second transition point $E_d$), the configurations below $E_c$, with label ``0'', remain structurally distinct from all configurations labeled ``1''. Consequently, the NN can still perfectly distinguish between configurations labeled ``0'' and ``1'', achieving an accuracy of 1.0 (in this case, the NN treats the two groups above $E_c$ as a single composite group). Similarly, the NN will achieve perfect accuracy for $E_c^\prime = E_d$, resulting in two intertwined ``W'' shapes in the $P(E_c^\prime)$ curve.

Finally, we emphasize that the theoretically predicted perfect ``W'' shapes may only be observed for phase transitions in large systems, as the system size approaches the thermodynamic limit. For finite systems such as polymers, which lack true phase transitions in the strict thermodynamic sense, structural distinctions between configurations close to a transition point are less pronounced. In such cases, we expect the sharpness of the ``W'' shapes to be reduced, as the classification accuracy will not reach a perfect 1.0 at the transition points. 

We would like to note that other unsupervised and semi-supervised machine learning approaches have been proposed for phase classification that do not require prior knowledge of the true phase labels. We selected the confusion scheme for its ability to directly identify all transition points within a given energy range, as well as for its simplicity and interpretability. One alternative NN-based method identifies a single transition point by training an NN on configurations from either end of a suspected transition region and locating the intersection of the NN outputs as a function of a thermodynamic variable~\cite{Carrasquilla2017np, Wei2017pre, Parker2022pre}. However, this method requires a pre-selected window assumed to contain a single transition and does not generalize well to systems with multiple transitions across a broader range. Moreover, for finite systems such as polymers that lack a sharp thermodynamic limit, the NN output may change gradually, making it difficult to pinpoint the transition reliably~\cite{Parker2022pre}. Another class of approaches involves dimensionality-reduction techniques such as principal component analysis, multidimensional scaling, or autoencoders. While these methods can produce low-dimensional representations of the configuration space in which different phases may be visually separable~\cite{Wang2016prb, Wang2017prb, Hu2017pre, Costa2017prb, Wetzel2017pre, Chng2018pre}, they do not inherently yield the transition points. In systems with sharp thermodynamic transitions, transition points may sometimes be inferred from the behavior of dominant principal components or latent variables as functions of temperature~\cite{Hu2017pre, Wetzel2017pre}. In contrast, for polymer systems, which are finite in nature and hence do not have a thermodynamic limit, such approaches are less effective.

\subsection{Neural-network architecture}

\begin{figure}
\includegraphics[width=.8\columnwidth, clip]{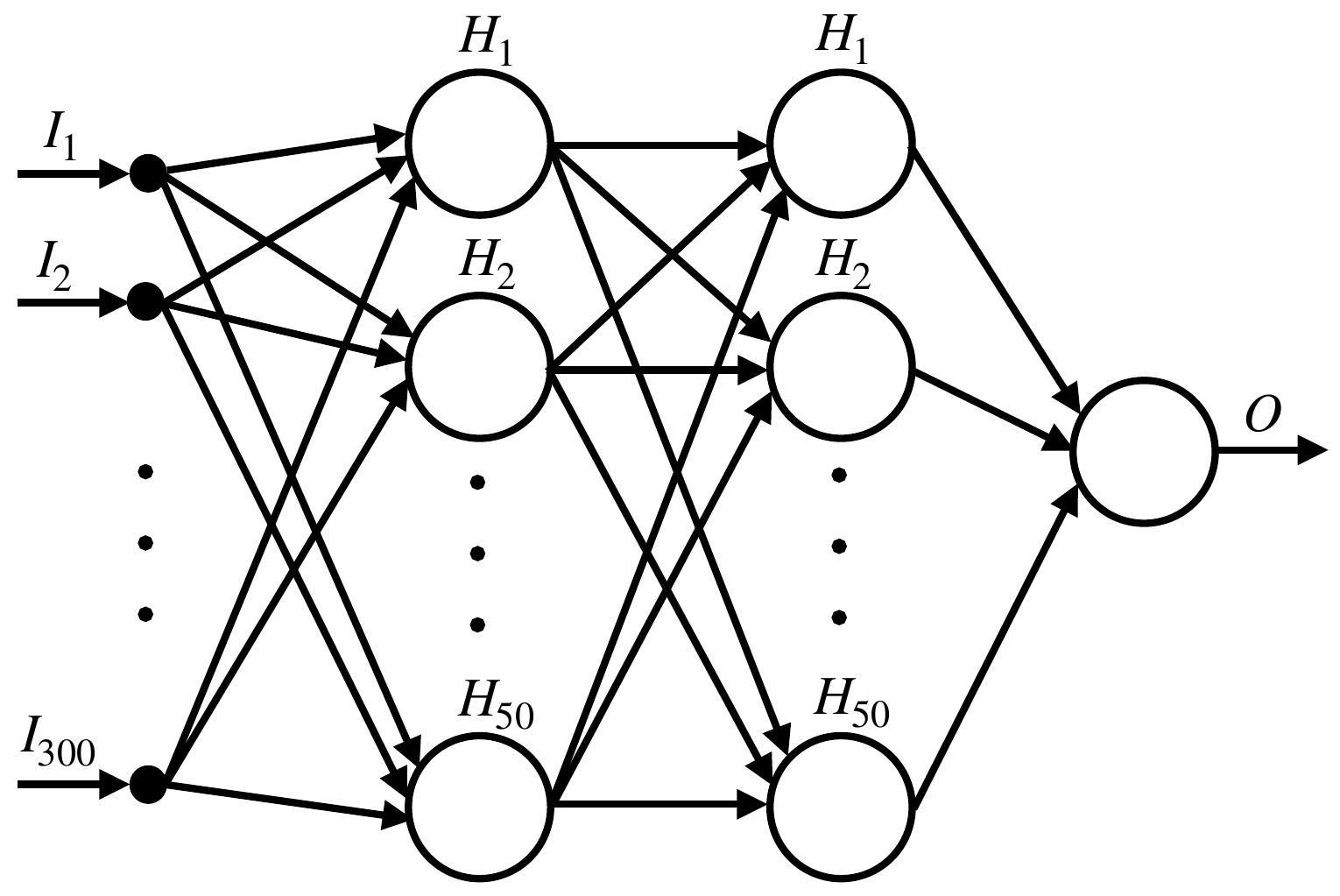}
\caption{Sketch of the neural-network architecture. We employ a fully-connected, feed-forward network with two hidden layers.\label{fig:nn_setup}}
\end{figure}
We employ a feed-forward NN architecture for the implementation of the confusion method. The input layer consists of 300 nodes, corresponding to the $3 \times N$ input features (with $N=100$) representing either the spatial coordinates of the monomers or the components of their dipole moments. This is followed by two hidden layers, each containing 50 neurons with ReLU activation functions. The output layer consists of a single neuron with a sigmoid activation function, whose output is interpreted as the probability that the input configuration belongs to the structural phase on the right side of the transition point. A sketch of the setup is shown in Fig.~\ref{fig:nn_setup}. The network was trained using the stochastic gradient descent (SGD) algorithm with a learning rate of 0.001 for 20 training epochs. Binary cross-entropy was used as the loss function, and no explicit regularization was applied. An example source code and dataset are available online~\cite{code_github}.
\section{Results} \label{sec:results}

In this work, we investigate structural transitions in the weak LJ attraction regime ($\eta < 0.1$), specifically focusing on $\eta = 0.02$ and $\eta = 0.06$.  
The equilibrium polymer configurations were generated from replica-exchange Wang--Landau runs as described in Sect.~\ref{sec:methods.rewl}. For each $\eta$ value, the confusion scheme was applied twice: first with the spatial coordinates of the monomers as input features for the NN, and then with the components of magnetic dipole moments as input features. Prior to feeding the input features to the NN, we preprocess the  configuration data to improve model performance. In the case of spatial coordinates, we subtract the center of mass of the polymer from the position coordinates of each monomer, which has been shown to yield improved results for polymers~\cite{Parker2022pre}. This translation reduces disparities in the length scales of position coordinates across different configurations, without introducing non-physical deformations to the polymer conformations. In the case of magnetic data, we normalize the dipole vectors by dividing by their magnitude~$|\vec{\mu}|$.
In each iteration of the confusion scheme, 70\% of the configurations were randomly selected for training, while the remaining 30\% were used for testing. To reduce statistical fluctuations, the confusion scheme was repeated 10 times, with different random selections of training and testing data in each repetition.

\begin{figure}[t!]
    \includegraphics[viewport=7 9 244 488, width=.9\columnwidth, clip]{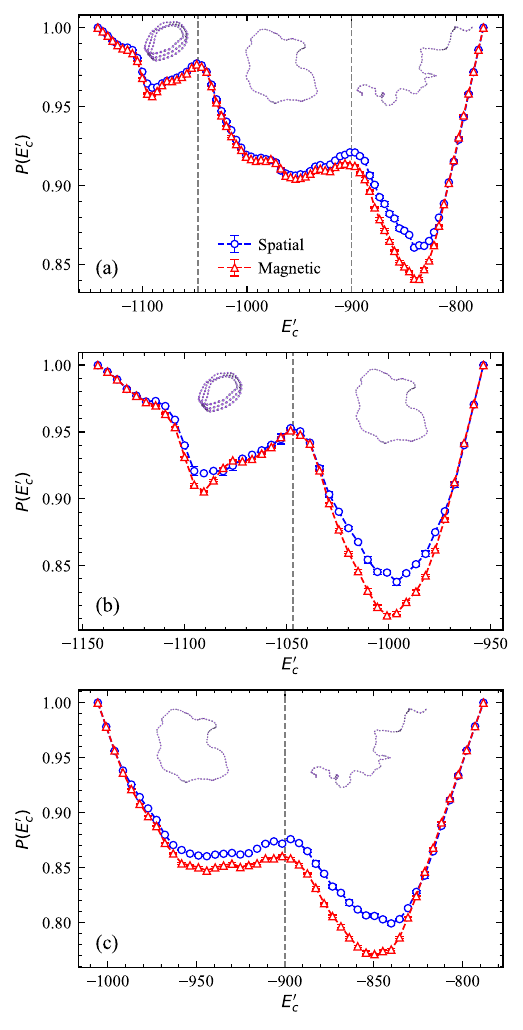}
    \caption{Test classification accuracy $P(E_c^\prime)$ as a function of the trial transition point $E_c^\prime$ for $\eta = 0.02$. Panels (a), (b), and (c) show results from the confusion method applied over different energy ranges: $E_c^\prime \in [-1143, -774]$, $E_c^\prime \in [-1143, -954]$, and $E_c^\prime \in [-1006, -788]$, respectively. Blue curves (circles) are from neural networks trained with the spatial coordinates of the monomers, while red curves (triangles) are from those trained with the magnetic dipoles. Vertical dashed lines indicate the approximate positions of the peaks in $P(E_c^\prime)$. Snapshots of typical polymer configurations observed in energy regions separated by the dashed lines are displayed.\vspace{-.38em}}
    \label{fig:confusion_eta_0.02}
\end{figure}

Fig.~\ref{fig:confusion_eta_0.02}\,(a) shows the test classification accuracy $P(E_c^\prime)$ as a function of the trial transition point $E_c^\prime$ for $\eta = 0.02$, obtained from the confusion scheme applied in the energy range $E_c^\prime \in [-1143, -774]$. The number of data points here corresponds to the number of collection bins in that energy range, as we move $E_c^\prime$ between bins for convenience. The results from the NN trained on spatial coordinates (blue curve with circles) qualitatively agree with those from the NN trained on magnetic dipoles (red curve with triangles). Each curve takes the form of two intertwined ``W'' shapes (though asymmetrical and slightly distorted), indicating the presence of two transitions at the peak positions $E_c^\prime \approx -1050$ and $E_c^\prime \approx -900$. At first glance, it may seem surprising that NNs trained on magnetic dipoles produce results that are qualitatively similar to those trained on spatial coordinates, especially considering that the predicted transitions correspond to structural transformations that alter the spatial distribution of monomers. However, it is important to note that the model constrains the orientation of each dipole to align with the line connecting the two anchoring points of the springs attached to the monomer. As a result, the dipole components indirectly encode information pertaining to the relative orientations of the bonds.

To demonstrate the robustness of the confusion scheme in detecting these transitions, we independently apply it over two shorter energy windows: one enclosing the transition at $E \approx -1050$ ($E_c^\prime \in [-1143, -954]$; Fig.~\ref{fig:confusion_eta_0.02}\,b), and the other enclosing the transition at $E \approx -900$ ($E_c^\prime \in [-1006, -788]$; Fig.~\ref{fig:confusion_eta_0.02}\,c). The resulting accuracy curves again follow the characteristic ``W'' shape, with peaks aligning with the transition points identified in Fig.~\ref{fig:confusion_eta_0.02}\,(a). These transitions can also be seen in the heat capacity shown in Fig.~\ref{fig:thermo}\,(top). The peak at $E \approx -900$ corresponds to the one at $T=0.82$ (Fig.~\ref{fig:thermo}\,top, inset) and had been identified before; see Fig.~11 in~\cite{2013-cerda}. The peak at $E \approx -1050$ corresponds to the temperature of $T=0.32$ and is located in the previously unexplored region of the phase diagram.\footnote{Since we obtain the density of states $g(E)$ from Wang--Landau runs and bin polymer configurations according to their energy, another natural choice to verify our results would be a microcanonical analysis~\cite{2006-junghans, 2011-schnabel, 2013-gai}. In fact, the inverse microcanonical temperature $1/T(E)$, calculated from the derivative of the logarithm of $g(E)$, shows the typical backbending at $E=-1048$ and $T(E)=0.325$. We do not show this analysis here, since it does not reveal any new information in this case.}
  
A first visual examination of polymer configurations reveals the structural differences among configurations across different regions of the energy space. At energies above $E = -900$, the polymers form open chains where magnetic dipoles locally align in a near head-to-tail formation, which becomes more pronounced as the energy decreases. Recall that a head-to-tail alignment of dipoles minimizes the magnetic dipole-dipole interaction given by Eq.~(\ref{dipdip}). At $E \approx -900$, the open chains transition into closed structures, with the two ends of the polymer attaching to one another, lowering the energy further due to the additional pair of aligned dipoles. These observations are consistent with the findings of Cerd\`{a} et al.~\cite{2013-cerda} for the weak LJ attraction regime ($\eta < 0.1$).    

Further decreasing the energy, the polymer undergoes a collapse transition into a compact helicoidal structure at $E \approx -1050$. This conformation forms through the twisting and folding of the single closed loop into multiple loops. These structures emerge as a consequence of the intricate interplay between the LJ attractive interaction and the magnetic dipole-dipole interaction~\cite{2013-cerda}. As energy decreases, a non-magnetic, flexible polymer would collapse into an isotropic globule, driven by the LJ attraction between monomers. However, in magnetic chains, the preference for head-to-tail alignment of dipoles effectively increases the stiffness of the chain, preventing complete collapse into a globule~\cite{2013-cerda, 2011-sanchez}. Instead, the polymer assumes a compact structure with locally aligned dipoles that strikes a balance between minimizing the LJ attraction and the dipole-dipole interaction. A previously presented phase diagram~\cite{2013-cerda} shows the emergence of helicoidal states only for $\eta \gtrsim 0.04$. Here we observe these states for all values of $\eta$, including $\eta = 0.00$, as our investigation extends into lower energy regions that were not explored previously.

\begin{figure*}[t]
    \includegraphics[viewport=7 9 418 402, width=0.8\textwidth, clip]{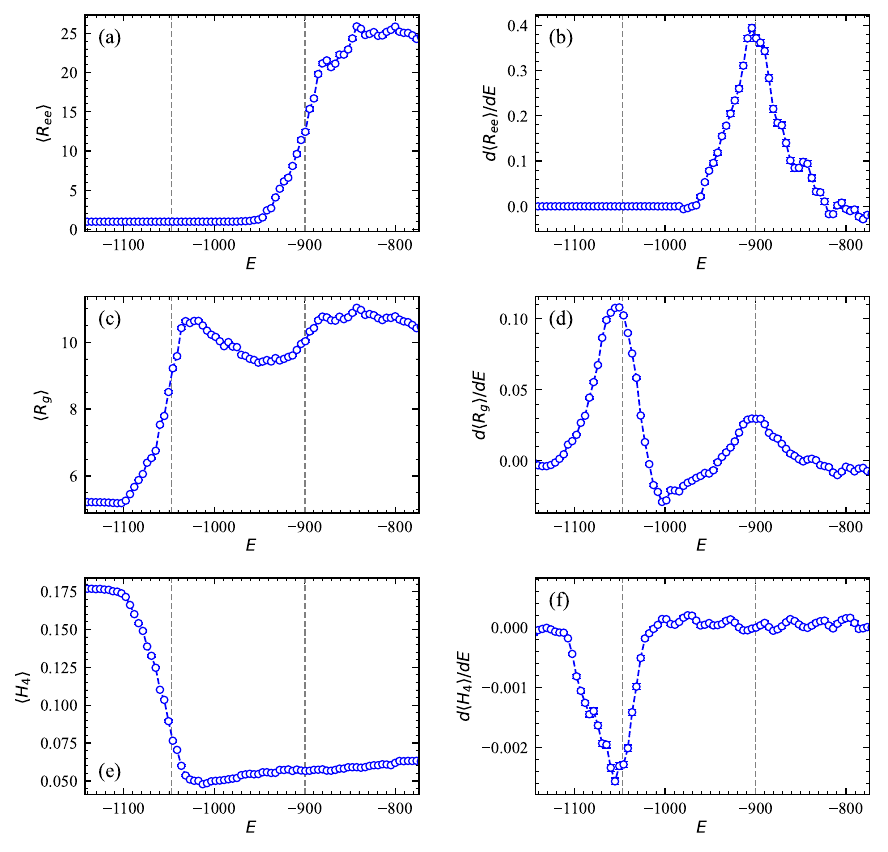}
    \caption{The average end-to-end distance $\langle R_{ee} \rangle$ (panel a), average radius of gyration $\langle R_g \rangle$ (panel c), and average helical order parameter $\langle H_4 \rangle$ (panel e) as functions of energy $E$ for $\eta = 0.02$. The corresponding derivatives with respect to energy are presented in panels (b), (d), and (f), respectively. Vertical, dashed lines correspond to the ones in Fig.~\ref{fig:confusion_eta_0.02}, indicating the positions of the peaks in the classification accuracy curves.\vspace{-.5em}}
    \label{fig:order_params_eta_0.02}
\end{figure*}

To further corroborate the results obtained through the confusion method, we calculate multiple observables typically used to investigate structural changes in polymers. The \textit{end-to-end distance} is defined as  
\begin{equation}  
    R_{ee} = |\vec{r}_1 - \vec{r}_N|,  
\end{equation}  
and is expected to be large for an extended, open chain but small for a collapsed or closed conformation.  

The elements of the \textit{gyration tensor} are given by  
\begin{equation}  
    R_{\alpha, \beta} = \frac{1}{2N^2} \sum_{i,j=1}^N \left(r_{i,\alpha} - r_{j, \alpha}\right) \left(r_{i,\beta} - r_{j,\beta}\right),  
\end{equation}  
where $\alpha$ and $\beta$ represent the Cartesian components $x$, $y$, and $z$.  
The eigenvalues $\lambda_1 \geq \lambda_2 \geq \lambda_3$ of the gyration tensor correspond to the principal axes of the polymer's spatial distribution and are used to compute various shape descriptors~\cite{aronovitz86jp,blavatska2010jcp,2013-arkın}. For example, the \textit{radius of gyration}, which quantifies the overall compactness of a structure, can be expressed as  
\begin{equation}  
    R_g = \sqrt{\lambda_1 + \lambda_2 + \lambda_3}.  
\end{equation}  

To characterize the global helical ordering associated with the onset of helicoidal structures, we calculate the $H_4$ parameter~\cite{2008-sabeur,2013-cerda},
\begin{equation}
    H_4 = \left| \frac{1}{N-2} \sum_{i=2}^{N-1} \left(\vec{r}_i - \vec{r}_{i-1}\right) \times \left(\vec{r}_{i+1} - \vec{r}_i\right) \right|.
\end{equation}
The $H_4$ parameter equals 1 for a perfect helix and 0 for an elongated, rod-like structure.

Fig.~\ref{fig:order_params_eta_0.02} shows the average end-to-end distance (a), average radius of gyration (c), and average $H_4$ parameter~(e) as functions of energy $E$ for the polymer configurations at $\eta = 0.02$. The corresponding derivatives with respect to energy are presented in panels (b), (d), and (f), respectively. The derivatives are computed using a five-point stencil after smoothing the data with the Savitzky-Golay filter. Vertical, dashed lines correspond to the ones in Fig.~\ref{fig:confusion_eta_0.02}. With decreasing energy, the average end-to-end distance rapidly decreases near the transition from open chains to simple closed loops ($E \approx -900$), after which it remains constant at the equilibrium value for a contact between two monomer, indicating a closed polymer. Its derivative exhibits a peak that very closely aligns with the transition point identified from the accuracy curves (Fig.~\ref{fig:confusion_eta_0.02}). Starting in the high-energy region, the average radius of gyration increases slightly with decreasing energy as the open chain stretches to reduce dipole-dipole interactions, as discussed earlier. It then decreases near the transition to a closed loop, before rising again as the loop stretches uniformly to adopt a more circular shape. Finally, it steeply declines near the transition to a helicoidal structure ($E \approx -1050$). The derivative of the radius of gyration exhibits two peaks, both closely aligning with the corresponding transition points. With decreasing energy, the average $H_4$ parameter shows a~rapid increase with the onset of helicoidal structures near $E \approx -1050$.  The corresponding derivative exhibits a dip that again closely aligns with the transition point identified through $P(E_c^\prime)$ vs.\ $E_c^\prime$ curves.

\begin{figure}[b]
    \includegraphics[viewport=7 9 244 326, width=\columnwidth, clip]{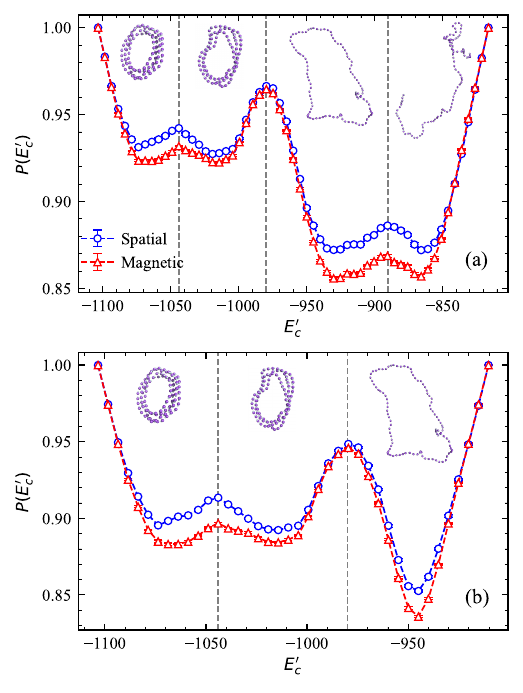}
    \caption{The test classification accuracy $P(E_c^\prime)$ as a function of the trial transition point $E_c^\prime$ for $\eta = 0.06$, obtained from the confusion method applied over the energy ranges $E_c^\prime \in [-1103, -816]$ (a) and $E_c^\prime \in [-1103, -910]$ (b). The blue curves (circles) correspond to neural networks trained on the spatial coordinates of the monomers, while the red curves (triangles) correspond to those trained on the components of magnetic dipole moments. Vertical dashed lines mark the locations of the peaks. Snapshots of typical polymer configurations observed in different energy regions are displayed above the curves.\vspace{-.4em}}
    \label{fig:confusion_eta_0.06}
\end{figure}

Figure~\ref{fig:confusion_eta_0.06}\,(a) shows the classification accuracy for $\eta = 0.06$ for the confusion scheme in the energy range $E_c^\prime \in [-1103, -816]$. The curves for both spatial and magnetic data exhibit a three-peak structure. The peaks at $E_c^\prime \approx -890$ and $E_c^\prime \approx -980$ indicate transitions from open chains to closed loops and from closed loops to helicoidal structures, respectively. They correspond to the peaks at $T=0.84$ and $T=0.58$ in Fig.~\ref{fig:thermo}\,(bottom). While the open-chain to closed-loop transition for $\eta = 0.06$ occurs at an energy and temperature very close to where it occurs for $\eta = 0.02$, the collapse transition to helicoidal states happens at a considerably higher temperature and energy. This is to be expected due to the increased LJ attraction strength for $\eta = 0.06$. The additional peak at $E_c^\prime \approx -1040$ suggests the presence of another transition, which was not observed for $\eta = 0.02$. Conformations with energies below this point are very similar in overall shape and structure to the helicoidal structures observed above this point, though they appear somewhat more symmetrical on average and tend to have a higher loop count overall. To confirm that this additional peak is not an artifact, the confusion scheme was repeated over a shorter energy range, $E_c^\prime \in [-1103, -910]$, which excludes the high-energy regime containing the open-chain to closed-loop transition. The resulting $P(E_c^\prime)$ curves (Fig.~\ref{fig:confusion_eta_0.06}\,b) exhibit a two-peak structure with peak positions that perfectly align with the first two peaks observed in panel (a).

\begin{figure*}
    \includegraphics[viewport=7 9 418 402, width=0.8\textwidth, clip]{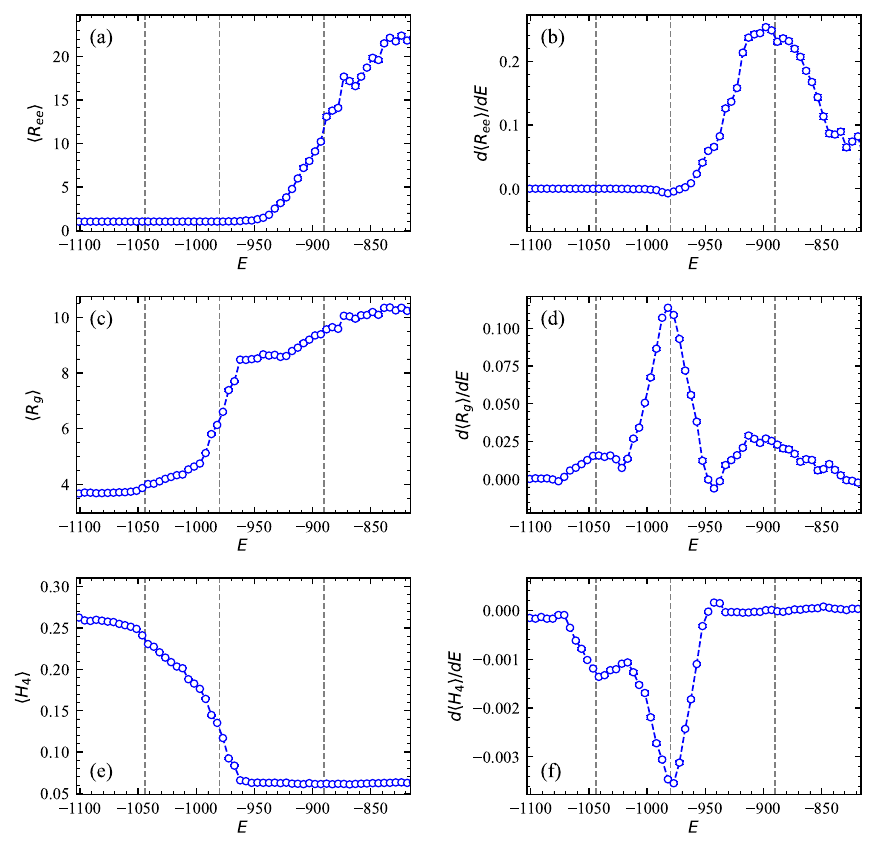}
    \caption{The average end-to-end distance $\langle R_{ee} \rangle$~(a), average radius of gyration $\langle R_g \rangle$~(c), and average helical order parameter $\langle H_4 \rangle$~(e) as functions of energy $E$ for $\eta = 0.06$. The corresponding derivatives with respect to energy are presented in panels (b), (d), and (f), respectively. Vertical dashed lines indicate the peak positions in the accuracy curves in Fig.~\ref{fig:confusion_eta_0.06}\,(a).\vspace{.3em}}
    \label{fig:order_params_eta_0.06}
\end{figure*}

The average end-to-end distance, radius of gyration, and $H_4$ parameter for $\eta = 0.06$, along with their derivatives, are shown in Fig.~\ref{fig:order_params_eta_0.06}. Similar to $\eta = 0.02$, these observables and their derivatives exhibit clear signals confirming the transitions from open chains to closed loops and from closed loops to helicoidal structures. In the vicinity of $E = -1040$, where the additional peak in $P(E_c^\prime)$ was observed, the derivative of the radius of gyration displays a subtle, broad feature resembling a peak, although it is much less pronounced than the peaks near $E \approx -980$ and $E \approx -890$. This leads to the speculation that, with decreasing energy, an abrupt reduction in the spatial dimensions of the helicoidal structures occurs at $E \approx -1040$. Furthermore, the derivative of the average $H_4$ parameter exhibits a small dip at $E \approx -1040$, suggesting an abrupt increase in the helicity of the structure with decreasing energy.

\begin{figure}
    \includegraphics[viewport=7 9 245 326, width=\columnwidth, clip]{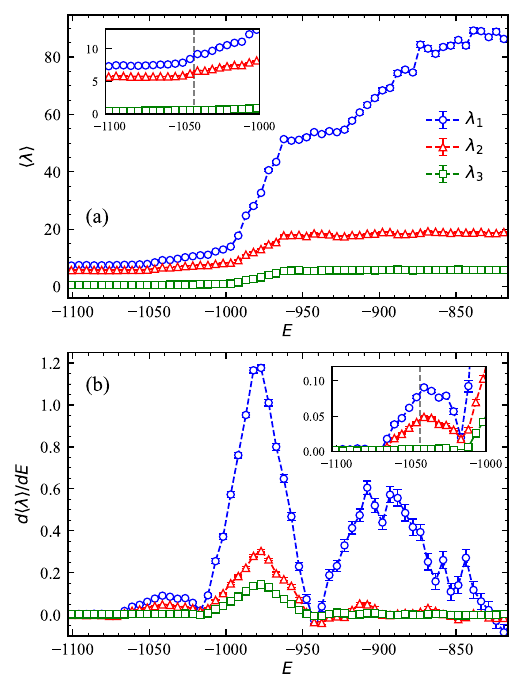}
    \caption{The average eigenvalues $\langle \lambda_1 \rangle$, $\langle \lambda_2 \rangle$, and $\langle \lambda_3 \rangle$ of the gyration tensor (a) and their derivatives with respect to energy (b) for $\eta = 0.06$. The insets provide a vertically enlarged view, highlighting finer details near $E \approx -1040$, where the additional peak in the $P(E_c^\prime)$ curve is observed (Fig.~\ref{fig:confusion_eta_0.06}). Vertical dashed lines in the insets mark the approximate position of this peak.}
    \label{fig:eigenvalues_eta_0.06}
\end{figure}

To gain deeper insight into the characteristics of helicoidal conformations and the structural transformation at low temperatures, we examine the eigenvalues of the gyration tensor, $\lambda_1$, $\lambda_2$, and $\lambda_3$. They quantify the spatial distribution of the monomers along the three principal axes and provide insight into the degree of asymmetry in a polymer conformation. Fig.~\ref{fig:eigenvalues_eta_0.06} shows the average eigenvalues and their derivatives with respect to energy. We note that $\langle \lambda_1 \rangle$ remains larger than $\langle \lambda_2 \rangle$ even in the energy regions corresponding to the single-loop and helicoidal phases, although their difference gradually diminishes with decreasing energy. This highlights the asymmetrical nature of single-loop and helicoidal structures, as confirmed by visual inspection of the conformations. Since the overall shape of helicoidal structures resemble cylinders with a small height-to-diameter ratio, the two most prominent principal directions (corresponding to the largest eigenvalues, $\lambda_1$ and $\lambda_2$) lie in a plane perpendicular to the cylindrical axis. Consequently, the observation that $\langle \lambda_1 \rangle$ is larger than $\langle \lambda_2 \rangle$ for helicoidal structures at moderate energies suggests that their cross-sectional regions are generally elongated, with a shape more akin to an ellipse than a circle. This aligns with the observations of Cerd\`{a} \textit{et al.}~\cite{2013-cerda}, who attributed such anisotropy in conformations to entropy maximization. 

While the two prominent peaks in the derivative of $\langle \lambda_1 \rangle$ correspond to the transitions from an open chain to a closed loop and then to a helicoidal structure, another much less pronounced, broad peak can be observed, which closely aligns with the peak in the $P(E_c^\prime)$ curve at $E \approx -1040$ (see the inset in Fig.~\ref{fig:eigenvalues_eta_0.06}\,b). This subtle peak appears in the derivatives of both $\langle \lambda_1 \rangle$ and $\langle \lambda_2 \rangle$ but is absent in the derivative of $\langle \lambda_3 \rangle$. This indicates an abrupt reduction in the cross-sectional dimensions of the helicoidal structures, while the ``heights'' of the structures remain practically unchanged. We speculate that this structural change likely results from an increase in the number of loops due to further twisting and folding of the helicoidal structure.  This interpretation is consistent with the abrupt change in the helicity of the structure observed at $E \approx -1040$, as indicated by the dip in the derivative of $\langle H_4 \rangle$ (Fig.~\ref{fig:order_params_eta_0.06}\,f). 

\begin{figure}
    \includegraphics[width=\columnwidth]{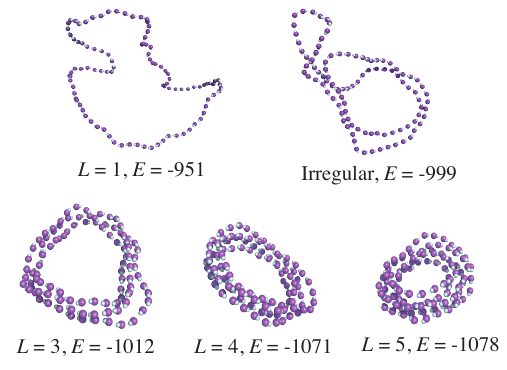}
    \caption{Configurations of magnetic Stockmayer polymers at $\eta=0.06$ with different loop counts $L$. Irregular structures, such as the one shown in the upper right corner, are primarily observed near the transition from closed-loop to helicoidal structures. The energy of each configuration ($E$) is indicated in the labels.}
    \label{fig:loop_structures_eta_0.06}
\end{figure}

To further test this hypothesis, we visually inspected numerous configurations to see how the number of loops~$L$ in helicoidal structures varies with energy over an energy range encompassing the transitions from single closed loops to helicoidal structures ($E \approx -980$) and then to more compact helicoidal structures ($E \approx -1040$). See Fig.~\ref{fig:loop_structures_eta_0.06} for some examples.
Within this energy range, we identify structures with loop counts $L = 1, 3, 4$, and $5$, as well as irregular structures that cannot be clearly classified as having a specific loop count. At energies above the transition from single closed loops to helicoidal structures ($E > -980$), single-loop conformations are dominant. Near this transition, we primarily observe irregular structures that resemble single-loop conformations in the process of folding into multi-loop structures. Interestingly, we do not observe clear instances of 2-loop structures. At energies between the two helicoidal transitions ($-1040 < E < -980$), 3-loop structures are the most prevalent, though 4-loop, 5-loop, and irregular structures are also observed. As the energy decreases further and approaching the transition at $E \approx -1040$, the number of 3-loop structures declines, while the more compact 4-loop and 5-loop structures become more frequent. The low-energy region below this transition point is largely dominated by 4-loop and 5-loop structures.

All results presented here are for polymers of length $N=100$ in order to be able to directly compare them with data previously obtained by Cerd\`{a} \textit{et al}.~\cite{2013-cerda}. Based on our results and previous experience, we conjecture that in the case of shorter polymers with $N\ll100$ some conformational phases will no longer exist: for very short polymer chains, helicoidal conformations are associated with very large energy and entropic penalties due to the high curvature. In such cases, adopting simple closed states is much more favorable. This has been observed in ferrofluids where magnetic particles can aggregate forming short chains and rings (closed states). Therefore, as $N$ decreases, helicoidal and partially collapsed regions in the conformational phase diagram are expected to occur at higher and higher values of $\eta$ while the region of closed states expands. In the limit of very short chains, we only expect regions associated with open, closed, and compact states. On the other hand, for $N \gg 100$ we believe that no qualitative changes will occur in the conformational phase diagram: $N=100$ chains are already quite representative of the long magnetic polymer regime.
\section{Summary} \label{sec:summary}
    
In this paper, we use a semi-supervised neural network based machine learning approach to explore structural phases and locate phase boundaries of a Stockmayer polymer, a chain of magnetic colloidal particles. A characteristic feature of the model is the presence of two competing interactions: short-range Lennard-Jones (LJ) attraction between the monomers and the magnetic dipole-dipole interactions. In the weak LJ attraction regime, characterized by $\eta < 0.1$, where $\eta$ quantifies the relative strength of the LJ attraction compared to the magnetic interaction, we observe three main types of polymer conformations: extended open chains at high energies, simple closed loops at moderate energies, and compact helicoidal structures at low energies. Our machine-learning approach successfully identifies all transition points between these structural phases. Moreover, our study extends into low-energy regions that were previously unexplored, in which we identify a new transition for $\eta = 0.06$ where the helicoidal structures collapse into more compact states with an increased number of loops.

A key hallmark of the confusion method is that, unlike conventional use cases of neural networks (NN), it does not require the true labels of the input training samples to be known in advance. Consequently, it can be applied over a chosen interval of energy (or equivalently, temperature, or a model parameter) to identify all transition points therein, without any prior knowledge of the existence (or absence) of transitions. These transitions will appear as peaks in the classification accuracy. For example, for $\eta=0.06$, we identify three peaks, corresponding to three structural transitions, through a single application of the confusion scheme. The NN architecture and the training algorithm were intentionally kept simple, and although NN training often requires careful fine-tuning of multiple hyperparameters (e.g., number of hidden layers, learning rate, etc.) to achieve satisfactory performance, we used \textit{ad hoc} hyperparameter choices and no further optimization, highlighting the robustness of the approach. We note, though, that the extent of required hyperparameter optimization and the effectiveness of the NN architecture may depend on the complexity of the system and the conformational phases.

The confusion method therefore provides a complementary perspective and is a valuable addition to existing tools to study structural transitions, such as micro\-canonical analysis~\cite{2006-junghans, 2011-schnabel}. It offers specific advantages and can provide a more comprehensive understanding of phase behavior in complex systems, in particular when compared to analyzing predetermined order parameters. Those must typically be known and chosen carefully to effectively capture different structural transitions, which often requires some human intuition~\cite{Vogel2010prl}. For example, observables derived from the gyration tensor, such as the radius of gyration and asymmetry or asphericity measures, have been used before~\cite{aronovitz86jp,blavatska2010jcp}. Here, the helical order parameter, $\langle H_4 \rangle$, was chosen specifically for its ability to characterize helicoidal states. The confusion scheme, in contrast, does not require any prior knowledge about the system or structural phases. The NN itself will internally construct its own ``order parameters'' to best categorize input configurations into distinct phases. Transition points can be directly extracted from the accuracy curve produced by the confusion scheme, while identifying them from order parameters often requires calculating numerical derivatives, which can be highly susceptible to noise in the data. For example, transitions between helicoidal states were barely discernible through the order parameters in this study, whereas the confusion scheme yielded a much clearer signal.

\section*{Acknowledgments} \label{sec:acknowledgments}

\noindent
All data were produced on the University of North Georgia's Pando computing cluster. JC acknowledges funding through grant PID2020-118317GB-I00 by MICIU/AEI/10.13039/501100011033. LA-UR-25-21917.

\bibliography{refs}

\end{document}